\documentclass[twocolumn,groupedaddress,aps,showpacs,pra]{revtex4}
\usepackage{multirow}
\usepackage{stmaryrd}
\usepackage{amssymb}
\usepackage{amsmath,amsthm}
\usepackage{amsfonts}
\usepackage{mathrsfs}
\usepackage{dsfont}
\usepackage{color}
\usepackage{epsfig,setspace,float}
\usepackage{bm}
\usepackage{setspace}
\input{Qcircuit}

\begin{document}
\newtheorem{proposition}{Proposition}
\newtheorem{prop}{Proposition}
\newtheorem{corollary}[proposition]{Corollary}
\newtheorem{definition}{Definition}
\newtheorem{theorem}{Theorem}
\newtheorem{lemma}{Lemma}

\theoremstyle{definition}
\newtheorem{example}{Example}

\title{Convex decomposition of dimension-altering quantum channels}

\author{Dong-Sheng Wang}
\affiliation{
Department of Physics and Astronomy, University of British Columbia, Vancouver, Canada}

\begin{abstract}
  Quantum channels, which are completely positive and trace preserving mappings,
  can alter the dimension of a system; e.g., a quantum channel from a qubit to a qutrit.
  We study the convex set properties of dimension-altering quantum channels,
  and particularly the channel decomposition problem in terms of convex sum of extreme channels.
  We provide various quantum circuit representations of extreme and generalized extreme channels,
  which can be employed in an optimization to approximately decompose an arbitrary channel.
  Numerical simulations of low-dimensional channels
  are performed to demonstrate our channel decomposition scheme.
\end{abstract}


\maketitle

\section{Introduction}
\label{sec:intr}

Quantum channels provide, arguably, the most general characterization
of quantum processes~\cite{Kra83}.
Unitary evolution, Markovian dynamics described by quantum master equation,
and positive operator-valued measure
are special cases of quantum channels.
Quantum channels also play vital roles in quantum computing and communication~\cite{NC00,AKN97},
and serve as a central subject for the tense study
such as quantum process tomography,
quantum simulation,
and quantum channel capacity.

Quantum channels are completely positive and trace preserving (CPTP) mappings~\cite{Sti55,Cho75},
and are usually considered as the dynamics on a qu$d$it,
i.e., a $d$-level quantum system $(d\geq 2)$.
Nevertheless, quantum channels can also serve as the mappings between two systems
with different dimensions;
e.g., a quantum channel from a qubit to a qutrit is possible without violating the CPTP condition.
We term channels that do not preserve dimension as dimension-altering (DA) quantum channels.
The DA quantum channels are more general than
dimension-preserving quantum channels and are important in many settings.
For instance, in quantum computing DA channels can be employed to drive a system towards a
decoherence-free subspace~\cite{LCW98}.
A matrix-product state~\cite{Sch05} can be viewed as the output state resulting from a sequence of DA channels
acting on the system and an ancilla.
Also, quantum open-system dynamics~\cite{Bre03} with particle creation and annihilation,
or particle gain and loss,
can be described by DA channels with the change of particle numbers
encoded as the change of dimension.

Our study in this work starts from the set of quantum channels instead of particular channels.
A primary property is that the set of quantum channels is convex.
This means any convex combination of channels still leads to a valid quantum channel.
The convex set of quantum channels and also other quantum objects such as states and observables
have been a major focus of mathematical characterization lately~\cite{AS08,HSZ14,HHP14,Pel14,FL13},
and
exploring the convexity can benefit tasks involving quantum channels,
such as quantum channel simulation~\cite{WBOS13,WS15}.

In this work we study the convex set properties of DA quantum channels.
In particular, we study the problem of channel decomposition
(also called partition) in terms of
convex sum of extreme channels~\cite{Cho75,RSW02,Rus07}.
The convex partition of quantum channels is fundamental for the study of channels.
Compared to the case of quantum states,
a mixed state can be partitioned into several pure states,
and the number of parts is lower bounded by the rank of the mixed state,
and the rank, or the logarithm of rank,
can be understood as the mixedness or noiseness of the mixed state,
which further relates to the entropy in the state.
An extreme quantum channel has a smaller rank than a general channel~\cite{Cho75}.
In physical terms,
a quantum process on a system may correspond to an extreme channel when
the system couples to a small environment,
whose dimension is no larger than that of the system.
The (logarithm of the) rank of a channel can be understood as the noiseness of it,
e.g., a rank-one channel is unitary and the degree of noise should be zero.
This observation also carries over to the infinite-dimensional case,
wherein a bosonic Gaussian quantum channel is extreme if and only if it has ``minimal noise''~\cite{Hol13}.

The extreme-channel decomposition is also motivated by Ruskai's conjecture~\cite{Rus07},
which states that a quantum channel from qu$n$its to qu$m$its can be decomposed as a convex sum
of $m$ ``generalized'' extreme channels,
each of which is at most of rank $n$.
The case for $n=m:=d$, i.e. qudit has been investigated recently~\cite{WS15},
which validates the conjecture numerically for low-dimensional cases:
qubit, qutrit, and two-qubit channels.
In this work we generalize the approach in Ref.~\cite{WS15}.
We provide the Kraus operator-sum and quantum circuit representations
of extreme and generalized extreme DA quantum channels,
which are then used in an optimization algorithm for channel decomposition.
The extreme quantum circuit construction for qudit case is generalized
to the DA extreme channels in different ways.
Also, numerical simulations for low-dimensional cases provide
support for Ruskai's conjecture in the more general setting.

In section~\ref{sec:pre} we present basic definitions of quantum channels and extreme channels,
and we study the simple yet nontrivial cases when the input or output system is of dimension one.
We study the construction of extreme DA quantum channels from qu$n$its to qu$m$its in section~\ref{sec:ext},
and provide illustrative examples of low-dimensional cases.
Also, we provide simpler constructions in section~\ref{sec:ext2},
and compare different constructions.
Next in section~\ref{sec:decom} we study the channel decomposition problem of arbitrary DA channels,
and present numerical simulations of low-dimensional cases.
We conclude in section~\ref{sec:conc}.

\section{Preliminaries}
\label{sec:pre}

\subsection{Notation}

The set of $n\times m$ complex matrices is denoted as $\mathcal{M}_{n,m}$,
and the set of $n\times n$ complex matrices is denoted as $\mathcal{M}_n$.
Tensor product is $\mathcal{M}_m\otimes \mathcal{M}_n= \mathcal{M}_m ( \mathcal{M}_n )$.
The set of density operators $\rho$ acting on an $n$-dimensional Hilbert space is $\mathcal{D}_n\subset \mathcal{M}_n$.
An identity matrix is denoted as $\mathds{1}$, and sometimes we use $\mathds{1}_n$
to indicate the dimension $n$.

Denote the set of CP mappings $\mathcal{E}: \mathcal{D}_n\rightarrow \mathcal{D}_m$ with Kraus operators $\{K_i\}$
and $\sum_i K_i K_i^\dagger =K$, $\sum_i K_i^\dagger K_i =L$ by $\mathscr{S}_{n,m}(K,L)$.
When there is no condition on the trace, we denote the set as $\mathscr{S}_{n,m}(K,\varnothing)$,
and when there is no condition on the unitality, we denote the set as $\mathscr{S}_{n,m}(\varnothing,L)$.
A CP trace preserving (CPTP) mapping is known as quantum channel.
Denote the set of channels $\mathcal{E}: \mathcal{D}_n\rightarrow \mathcal{D}_m$
as $\mathscr{S}_{n,m}$.
If $n=m:=d$, the set is denoted as $\mathscr{S}_d$ for simplicity.
We use the term $(n,m)$-channel $\mathcal{E}$ to indicate that
$\mathcal{E}: \mathcal{D}_n\rightarrow \mathcal{D}_m$.
We denote the transpose operation by a superscript $t$;
e.g. the transpose of a matrix $A$ is denoted by $A^t$.
If $\mathcal{E}(\rho)=\sum_i K_i \rho K_i^\dagger$, the transpose of $\mathcal{E}$
is denoted as $\mathcal{E}^t$ with $\mathcal{E}^t(\rho)=\sum_i K_i^\dagger \rho K_i$.
A unital channel is CPTP and identity-preserving, i.e. $\mathcal{E}^t(\mathds{1})=\mathcal{E}(\mathds{1})=\mathds{1}$.

\subsection{Representations of quantum channels}

We first recall the basic representations of quantum channels below,
including the Kraus operator-sum representation,
Choi state representation,
and unitary evolution representation.

A linear mapping $\mathcal{E}: \mathcal{D}_n \rightarrow \mathcal{D}_m$
is completely positive iff it takes the form~\cite{Cho75}
  \begin{equation}\label{eq:channeldef}
    \mathcal{E}(\rho)=\sum_i K_i \rho K_i^\dagger,
  \end{equation}
for all $\rho \in \mathcal{D}_n$, and $K_i\in \mathcal{M}_{m,n}$.
When $\sum_i K_i^\dagger K_i=\mathds{1}$, the map is trace-preserving, and known as quantum channel.
Given a channel $\mathcal{E}$, there exists a canonical representation~\cite{Cho75,NC00} by a set $\{K_i\}$,
which is linearly independent,
and the cardinality of this set is called the {\em Kraus rank},
denoted as $r_\mathcal{E}$, and $r_\mathcal{E}\leq mn$.

The Choi-Jamio{\l}kowski isomorphism~\cite{Jam72,Cho75}
$\mathcal{J}:\mathscr{S}_{n,m}\rightarrow \mathcal{D}_{nm}$
maps a quantum channel $\mathcal{E}\in \mathscr{S}_{n,m}$ into a quantum state,
known as Choi state
\begin{equation}\label{eq:choi}
\mathcal{C}: =  \mathcal{E}\otimes\mathds{1} (|\eta\rangle\langle\eta|) \in \mathcal{D}_{nm}\subset \mathcal{M}_m\otimes \mathcal{M}_n,
\end{equation}
with bipartite maximally entangled state
\begin{equation}\label{}
|\eta\rangle=\frac{1}{\sqrt{n}}\sum_{i=0}^{n-1}|i,i\rangle.
\end{equation}
The channel $\mathcal{E}$ and the state $\mathcal{C}$
are known as the dual of each other.
It is clear that the dual channel of $|\eta\rangle$ is the identity channel.
The condition of complete positivity is equivalent to
the positive semidefiniteness of the Choi state; i.e., $\mathcal{C}\geq 0$.
The rank of $\mathcal{C}$, denoted as $r_\mathcal{C}$, equals the
Kraus rank of the channel: $r_\mathcal{C}=r_\mathcal{E}$.

Physically, given a system \textsc{s} with state $\rho$,
a quantum channel $\mathcal{E}$ acting on it can be realized
by the interaction specified by a unitary evolution $U$,
with one ancillary system \textsc{a} prepared in state $\sigma$
such that
\begin{equation}\label{eq:cpunitaryrep}
  \mathcal{E}: \mathcal{D}_n \rightarrow \mathcal{D}_m:
  \rho \mapsto \mathcal{E}(\rho)=\text{tr}_\textsc{a'} U(\sigma\otimes\rho) U^\dagger,
\end{equation}
for \textsc{a'} as the output ancilla.
When $r_\mathcal{E}=mn$, $U\in SU(m^2n)$.
That is, the input include the $n$-dimensional system and an $m^2$-dimensional ancilla,
and the output include an $m$-dimensional system and $mn$-dimensional ancilla.
Usually, the ancilla state $\sigma$ is chosen as the ground state $|0\rangle\langle 0|$,
and then the Kraus operators take the form $K_i=\langle i|U|0\rangle$
such that projection on the output ancilla realizes the Kraus operators~\cite{NC00}.

\begin{figure}[t!]
  \centering
  \includegraphics[width=.15\textwidth]{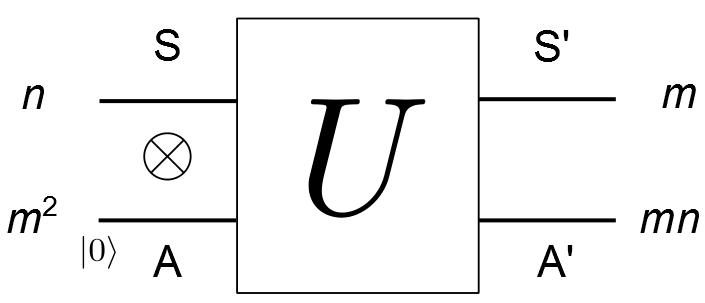}
  \includegraphics[width=.15\textwidth]{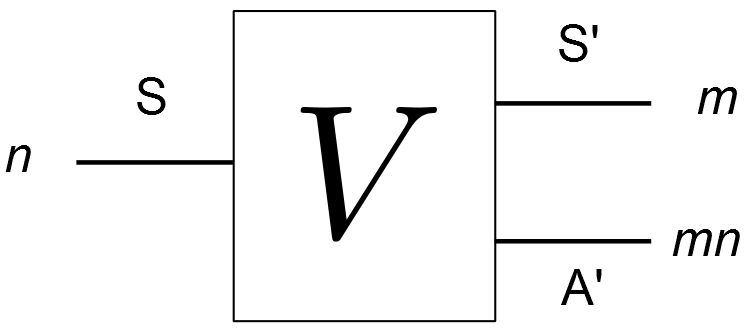}
  \includegraphics[width=.15\textwidth]{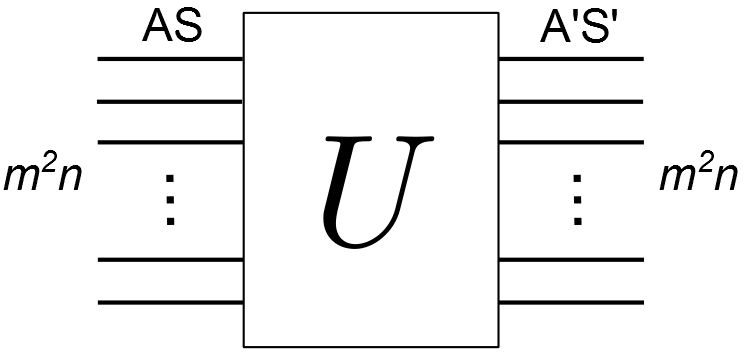}
  \caption{The graphic representations of $U$ (left), $V$ (middle),
  and $U$ in the unary form (right).}\label{fig:UV}
\end{figure}

\noindent The graphic representation of $U$ is shown in Fig.~\ref{fig:UV} (left).
In the circuit graph each wire represents a physical system,
\textsc{s} (\textsc{s}') for the input (output) system,
and \textsc{a} (\textsc{a}') for the input (output) ancilla.
The symbol $\otimes$ stands for the tensor product of the
two spaces leading to a total dimension $m^2n$.
The input ancilla \textsc{a} is at state $|0\rangle$
while the final ancilla \textsc{a}' is traced out.

Actually, the implementation of a quantum channel $\mathcal{E}$
does not require a complete implementation of $U$,
since the state of the input ancilla is fixed.
This is clear from the matrix form
\begin{equation}\label{}
  U=\begin{pmatrix}
    K_0 & \cdot & \cdots & \cdot \\
    K_1 & \cdot & \cdots & \cdot \\
    \vdots & \vdots & \cdots & \vdots
  \end{pmatrix},
\end{equation}
if $U$ is expanded in the basis of $\textsc{a}\otimes\textsc{s}$ for the input space,
and $\textsc{a}'\otimes\textsc{s}'$ for the output space,
wherein the Kraus operators only occupy the first block-column of it.
The first block-column forms an isometry $V:=U|0\rangle$
for $|0\rangle$ the input ancilla state
such that $V^\dagger V=\mathds{1}$ while $V V^\dagger \neq \mathds{1}$.
That is, the dynamics of a quantum channel can be equivalently expressed in the following three ways
\begin{align}\label{} \nonumber
  \mathcal{E}(\rho) &= \text{tr}_\text{A'} U (|0\rangle\langle 0|\otimes \rho)U^\dagger \\ \nonumber
   &= \sum_i K_i \rho K_i^\dagger \\
   &=\text{tr}_\text{A'} V \rho V^\dagger.
\end{align}

The graphic representation of $V$ is shown in Fig.~\ref{fig:UV} (middle).
Furthermore, for the understanding of the dimension-altering effect,
it would be more suitable to use a direct-sum structure of space
instead of a tensor-product structure employed above.
That is, we use the circuit graph shown in Fig.~\ref{fig:UV} (right)
to represent the implementation of $U$.
This is a unary representation,
i.e.\, each wire has dimension one and the direct sum
of all the wires results in the whole space with dimension $m^2n$.
The merit of this direct-sum representation is that
one does not have to consider the underlying, if there is,
tensor-product structure,
and it is suitable for analysis of the input and output information
separately.
Sometimes we also use a hybrid representation,
wherein we use unary wires for each subsystem while
there is still a tensor-product structure between subsystems.

\subsection{Convex set of quantum channels}

The set $\mathscr{S}_{n,m}$ is convex,
which means that $p\mathcal{E}_1+(1-p)\mathcal{E}_2\in \mathscr{S}_{n,m}$
for $p\in (0,1)$ and $\mathcal{E}_1, \mathcal{E}_2\in \mathscr{S}_{n,m}$.
A channel that cannot be written as a nontrivial convex sum of any other channels
is extreme.
An elegant characterization of extreme channels is provided by Choi~\cite{Cho75}.
Note that the original theorem of Choi also applies for trace non-preserving cases,
while here we limit to CPTP cases.
\begin{theorem}[Choi's theorem~\cite{Cho75}]
\label{the:Choiext}
Let $\mathcal{E}: \mathcal{D}_n\rightarrow \mathcal{D}_m$
with its canonical representation $\{K_i\}$.
Then $\mathcal{E}$ is extreme in $\mathscr{S}_{n,m}$
iff $\{K_i^\dagger K_j\}$ is linearly independent.
\end{theorem}
From this theorem it is easy to obtain that the rank of an extreme channel
is upper bounded by $n$.
We note that the set $\mathscr{S}_{n,m}$ is very different from the set of density operators,
for which case the extreme points are all of rank one,
i.e. pure states.

The collection of extreme channels also form a set, yet not closed.
Define the set of rank up to $n$ channels as $\mathscr{S}^{\leq n}_{n,m}$.
A primary property of $\mathscr{S}^{\leq n}_{n,m}$ is stated by the following theorem.
\begin{theorem}[Ruskai's theorem~\cite{Rus07}]
  The set $\mathscr{S}^{\leq n}_{n,m}$ is the closure of the set of extreme channels.
\end{theorem}
A channel $\mathcal{E}\in \mathscr{S}^{\leq n}_{n,m}$ is called a generalized extreme channel,
which may be extreme if the linear independence condition is satisfied,
quasi-extreme if not~\cite{Rus07}.
It is also shown recently, from a semi-algebraic geometry approach,
that the set of extreme channels dominates the set of generalized extreme channels~\cite{FL13}.
For clarity,
we denote an extreme channel as $\mathcal{E}^\text{e}$,
a generalized extreme channel as $\mathcal{E}^\text{g}$,
and a quasi-extreme channel as $\mathcal{E}^\text{q}$.
It is clear that a channel $\mathcal{E}^\text{q}$ can be written as a convex sum
of at least two extreme channels for its decomposition.

\subsection{Extreme channels in the set $\mathscr{S}_{n,1}$ or $\mathscr{S}_{1,m}$}

We find it is beneficial to start from the most basic cases,
the sets $\mathscr{S}_{n,1}$ and $\mathscr{S}_{1,m}$,
for which properties have been studied in literature, e.g., Ref~\cite{CDP08b},
while here we analyze them from the perspective of extreme channels.

\subsubsection{The set $\mathscr{S}_{n,1}$}

A quantum channel
$\mathscr{S}_{n,1}\ni \mathcal{E}: \mathcal{D}_n \rightarrow \mathbb{C}$
basically maps a quantum state $\rho \in \mathcal{D}_n$ into a null state.
This can be interpreted as that the system is lost or annihilated,
or just ignored.
We start from the case $n=2$,
which is simple yet can be generalized directly.

For $n=2$, the rank of an arbitrary channel is upper bounded by two.
Suppose the rank is one, and let the single Kraus operator be $K:=(a, b)$,
which has to satisfy the normalization condition $K^\dagger K=\mathds{1}_2$.
It is easy to observe that it is not possible to find $a$ and $b$ to satisfy
the normalization, which means that there is no rank-one channel in $\mathscr{S}_{2,1}$.
The only possible channels are then rank-two channels.
Let two Kraus operators be $K_0:=(a, b)$ and $K_1:=(c, d)$,
then normalization requires $a^2+c^2=1$, $b^2+d^2=1$, and $a^*b+c^*d=0$.
Define a matrix $R=\begin{pmatrix}   a & b \\ c & d \end{pmatrix},$
which is unitary.
This implies that there exist two new Kraus operators $M_0=(1, 0)$, and $M_1=(0, 1)$
such that $K_0=M_0 R$, $K_1=M_1 R$.
Now the matrix $R$ can be interpreted as a rotation on the system
before the channel defined by $M_0$ and $M_1$,
which is actually the trace operation.
This analysis reveals that, up to a basis transformation,
any rank-two channel in $\mathscr{S}_{2,1}$ is equivalent to the trace operation.
Also, the set $\{M_0^\dagger M_0, M_0^\dagger M_1, M_1^\dagger M_0, M_1^\dagger M_1\}$
is linearly independent,
which means the trace operation is an extreme channel.
We then conclude that the set $\mathscr{S}_{2,1}$ only contains a single point,
which is the trace operation.
In terms of Choi state,
we find that the Choi state of trace operation is the identity operator.

The analysis generalizes to all $n$, and we find that $\mathscr{S}_{n,1}$ is just a single point,
which is the trace operation.
Furthermore, when applying to a part of a composite system
it corresponds to the partial trace operation.

\subsubsection{The set $\mathscr{S}_{1,m}$}

We then study the case of $\mathscr{S}_{1,m}$,
which is quite different from the set $\mathscr{S}_{n,1}$.
As the normalization of Kraus operators is one, i.e. $\mathds{1}_1$,
applying Choi's theorem leads to that
extreme channels are all rank one.
This means the Choi state of an extreme channel is pure.
As a result,
an arbitrary channel with rank $2\leq r\leq m$ is a convex sum of $r$
extreme channels.
There is no quasi-extreme channel,
which means the set of extreme channels is closed.

In terms of Choi state, it turns out the set $\mathscr{S}_{1,m}$
is just the set of density operators $\mathcal{D}_m$,
for which the pure states are the extreme points,
while all mixed states are convex sum of pure states.
In other words,
a quantum state can be viewed as a quantum channel,
which creates or prepare a quantum state from a null state.

\section{Extreme channel circuit from cosine-sine decomposition}
\label{sec:ext}

In this section we study the general construction of generalized extreme channels
based on cosine-sine decomposition (CSD)~\cite{PW94} of unitary matrices.

For a general unitary operator $U\in SU(N)$ and any $2\times 2$ partitioning
\begin{equation}\label{eq:Upart}
  U=\begin{pmatrix}
    U_{11} & U_{12} \\ U_{21} & U_{22}
  \end{pmatrix},
\end{equation}
for $U_{11}\in \mathcal{M}_{r_1,c_1}$,
$U_{12}\in \mathcal{M}_{r_1,c_2}$,
$U_{21}\in \mathcal{M}_{r_2,c_1}$,
$U_{22}\in \mathcal{M}_{r_2,c_2}$,
such that $r_1+r_2=c_1+c_2=N$,
the CSD is stated as follows.
\begin{theorem}[The CSD~\cite{PW94}]
  For unitary operator $U$ in Eq.~(\ref{eq:Upart}) there exist unitary operators
  $W_1\in SU(r_1)$, $V_1\in SU(c_1)$, $W_2\in SU(r_2)$, and $V_2\in SU(c_2)$
  such that $U=WMV$ for $W=\emph{diag}(W_1,W_2)$, $V=\emph{diag}(V_1,V_2)$,
  \begin{equation}\label{}
    M=\left(\begin{array}{ccc|ccc}
      \mathds{1} & & & O & & \\
      & C & & & -S & \\
      & & O & & & \mathds{1} \\ \hline
      O & & & \mathds{1} & & \\
      & S & & & C & \\
      & & \mathds{1} & & & O
    \end{array}\right),
  \end{equation}
  and for $C\equiv \emph{diag}(\cos\theta_1,\dots,\cos\theta_s)$,
  $S\equiv \emph{diag}(\sin\theta_1,\dots,\sin\theta_s)$
  such that $C^2+S^2=\mathds{1}$.
  In the matrix of $M$ the big $O$ are zero matrices and $\mathds{1}$ are identity matrices and,
  depending on $U$ and the partition,
  may have no rows or columns, or be nonexistent.
\end{theorem}
The forms of the big $O$s and $\mathds{1}$s in the theorem will be clear
for explicit examples.
In the following we will study the quantum circuits for
generalized extreme qudit channels for $d=2,3,4$,
and $(n,m)$-channels for the cases of
$(2,3)$-, $(3,2)$-, $(2,4)$-, and $(4,2)$-channels.

\subsection{Quantum circuit of generalized extreme qudit channels}
\label{sec:cirgextd}

\subsubsection{Extreme qubit channel}

A generalized extreme qubit channel can be represented by two Kraus operators $K_0$
and $K_1$, and the dilated unitary operator $U\in SU(4)$ can be partitioned as
 \begin{equation}\label{eq:Uqubgext}
  U=\begin{pmatrix}
    K_0 & F_0 \\ K_1 & F_1
  \end{pmatrix},
\end{equation}
wherein $F_0$ and $F_1$ are matrices such that $U$ is unitary.
Our purpose is to find a general quantum circuit representation of
arbitrary generalized extreme qubit channels
from the CSD of $U$.
From
\begin{equation}\label{eq:Uqubgext1}
  U=\begin{pmatrix}     W_1 & \\ & W_2   \end{pmatrix}
  \begin{pmatrix}     C & -S \\ S & C   \end{pmatrix}
  \begin{pmatrix}     V_1 & \\ & V_2   \end{pmatrix}:=WMV,
\end{equation}
for $C\equiv \text{diag}(\cos\theta_1,\cos\theta_2)$,
$S\equiv \text{diag}(\sin\theta_1,\sin\theta_2)$,
$V_1,V_2,W_1,W_2\in SU(2)$,
we find the isometry
\begin{equation}\label{}
  U|0\rangle=\begin{pmatrix}     W_1 & \\ & W_2   \end{pmatrix}
  \begin{pmatrix}     C  \\ S   \end{pmatrix}
  V_1,
\end{equation}
and the two Kraus operators are
\begin{equation}\label{eq:Kqubgext}
K_0=W_1 C V_1, \; K_1=W_2 S V_1.
\end{equation}
The implementation of $K_0$ and $K_1$ does not depend on $V_2$,
since the input ancilla state is fixed as $|0\rangle$.
The representation~(\ref{eq:Kqubgext}) generalizes
the form in Ref.~\cite{WBOS13}
in that the posterior rotations for the two Kraus operators
are different arbitrary rotations.

A quantum circuit diagram of an arbitrary generalized extreme qubit channel is shown below
\[ \Qcircuit @C=.5em @R=.7em @!R {
\lstick{\rho_{\text{in}}} & \gate{V}      & \ctrlo{1}  & \qw  & \ctrl{1} &\qw                   &\gate{W_i} & \rstick{\rho_{\text{out}}} \qw     \\
\lstick{|0\rangle}               & \qw & \gate{R_y(2\theta_1)} & \qw    &\gate{R_y(2\theta_2)}      & \meter & \control \cw \cwx                                & \rstick{i} \cw
}\]

\noindent where we have used $V$ to denote $V_1$,
and the posterior rotations $W_i$ ($i=0,1$)
are classically controlled by the projection on the output ancilla.
The two controlled-rotations in the middle is a multiplexer~\cite{MV06}
to realize $M$ in Eq.~(\ref{eq:Uqubgext1}).
Furthermore, one controlled-NOT gate can be saved for the multiplexer
using circuit equivalence for qubit gates~\cite{WBOS13}.

We also employ unary representation and hybrid representation of quantum circuit,
which is especially suitable for DA quantum channels.
The hybrid representation of the generalized extreme qubit channel circuit above is
\[  \includegraphics[width=.35\textwidth]{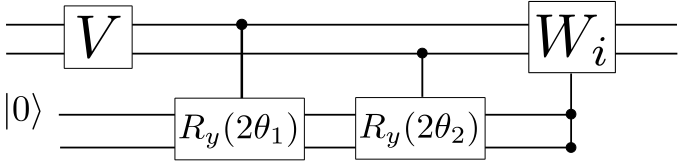}
\]

\noindent There is still a tensor-product structure $\textsc{s}\otimes\textsc{a}$
for the system and ancilla,
while for each of them the two wires are of direct-sum structure.
Note a control in the unary circuit is represented by a filled dot
together with a vertical line towards the controlled rotation.
The trace operation on the ancilla is represented by a vertical line connecting its two wires,
and the classical control is represented by filled dots together with a vertical line towards the controlled rotation.

\subsubsection{Extreme qutrit channel}

A generalized extreme qutrit channel can be represented by three Kraus operators $K_0$, $K_1$,
and $K_2$, and the dilated unitary operator $U\in SU(9)$ can be partitioned as
 \begin{equation}\label{eq:Uqutgext}
  U=\begin{pmatrix}
    K_0 & F_0 & E_0 \\ K_1 & F_1 & E_1 \\ K_2 & F_2 & E_2
  \end{pmatrix},
\end{equation}
whereas $F_0$, $F_1$, $F_2$, $E_0$, $E_1$, and $E_2$ are matrices such that $U$ is unitary.
By employing CSD twice~\cite{KP06} it holds that $ U=ABC G DEF$
for
\begin{align}\label{} \nonumber
A&=\text{diag}(W_1,W_2,W_3), & D&=\text{diag}(V_1,V_2,V_3), \\ \nonumber
B&=\text{diag}(\mathds{1},M_1), & E&=\text{diag}(\mathds{1},M_2), \\ \nonumber
C&=\text{diag}(\mathds{1},Z_1,Z_2),  & F&=\text{diag}(\mathds{1},Y_1,Y_2), \\
G&=\text{diag}(M_0,\mathds{1}),
\end{align}
for $W_1,W_2,W_3\in SU(3)$, for $Z_1,Z_2\in SU(3)$, for $V_1,V_2,V_3\in SU(3)$,
 for $Y_1,Y_2\in SU(3)$
with
\begin{equation}\label{eq:qutMulti}
  M_i=\begin{pmatrix}
    C_i & -S_i \\ S_i & C_i
  \end{pmatrix}, \; i=0,1,2,
\end{equation}
and $3\times 3$ diagonal matrices $C_i$ and $S_i$
such that $C_i^2+S_i^2=\mathds{1}$.

We find
\begin{align}\label{eq:Kqutgext}\nonumber
  W_1 C_0 V_1 & = K_0 ,\\ \nonumber
  W_2 C_1 Z_1 S_0 V_1 & = K_1,  \\
  W_3 S_1 Z_1 S_0 V_1 & = K_2.
\end{align}

Compared with the qubit case~(\ref{eq:Kqubgext}),
we see that $K_0$ takes a similar form,
while $K_1$ and $K_2$ are generalized nontrivially
due to the presence of a unitary operator $Z_1$ (denoted as $Z$ for simplicity).
Also the implementation of the qutrit multiplexer,
which is in general called a uniformly controlled rotation,
or multivalued multiplexer~\cite{MV06},
would be similar with the qubit case.

The hybrid representation of the qutrit circuit is
\[
  \includegraphics[width=.4\textwidth]{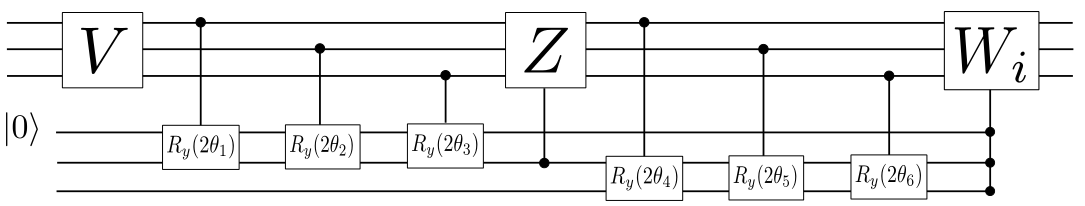}
\]

In order to have a simpler circuit diagram,
we employ the circuit diagram for a multivalued multiplexer~\cite{MV06}.
A Givens rotation~\cite{NC00}, which is a two-level unitary operator,
nontrivially acting on a subspace spanned by basis states $|i\rangle$
and $|j\rangle$ takes the form
\begin{equation}\label{eq:Givens}
  G_{ij}(\theta):= \cos\theta (|i\rangle\langle i| + |j\rangle\langle j|) +
                    \sin\theta (|j\rangle\langle i| - |i\rangle\langle j|).
\end{equation}
Note that a Givens rotation $G_{ij}(\theta)$ just acts as $R_y(2\theta)$
on the subspace spanned by $|i\rangle$ and $|j\rangle$.
We also define a controlled-Givens rotation as
\begin{equation}\label{eq:cGivens}
  C_l G_{ij}(\theta):= |l\rangle\langle l|\otimes G_{ij}(\theta).
\end{equation}
A multivalued multiplexer $M_{ij}$ takes the form
\begin{equation}\label{eq:quditMulti}
M_{ij}:=\prod_l C_l G_{ij}(\theta_l).
\end{equation}
A multivalued multiplexer is also called a uniformly controlled rotation,
due to the fact that the controller runs over all basis states.
As an application, the qutrit multiplexer $M_0$ in Eq.~(\ref{eq:qutMulti})
can be decomposed as $M_0=C_0 G_{01}(\theta_0) C_1 G_{01}(\theta_1) C_2 G_{01}(\theta_2)$.
Note that our multivalued multiplexer has a restriction on the target subspace,
while more general multivalued multiplexer can also be defined without such restriction.

A circuit diagram for the multiplexer $M_{ij}$ is
\[
  \includegraphics[width=.15\textwidth]{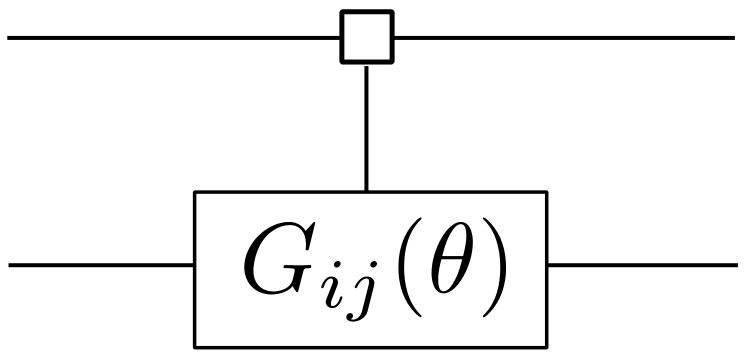}
\]

\noindent Here each wire represents a qudit and there is a tensor-product structure.
Using the notation of multivalued multiplexer,
we can simplify the quantum circuit diagram of an arbitrary generalized extreme qutrit channel as
\[  \includegraphics[width=.35\textwidth]{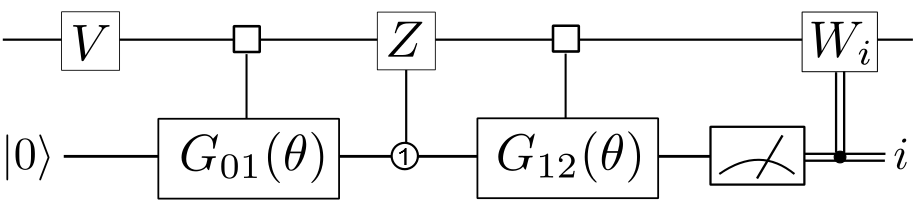}\]

\noindent wherein \textcircled{1} means a control when the ancilla is at state $|1\rangle$,
while in general \textcircled{\emph{i}} means a control when the controller is at state $|i\rangle$.

\subsubsection{Extreme two-qubit channel}

Following a similar procedure,
for generalized extreme two-qubit channels we find
\begin{align}\label{eq:Kqubbgext}\nonumber
                  W_1 C_0 V & = K_0, \\ \nonumber
            W_2 C_2 Y S_0 V & = K_1,  \\ \nonumber
  W_3 C_1 Z S_2 Y S_0 V & = K_2,  \\
  W_4 S_1 Z S_2 Y S_0 V & = K_3.
\end{align}
The quantum circuit diagram of an arbitrary generalized extreme two-qubit channel is
\[
  \includegraphics[width=.4\textwidth]{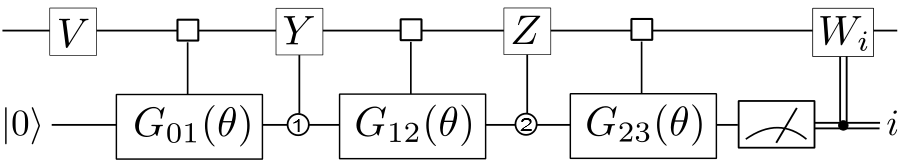}
  \]

\noindent for arbitrary two-qubit unitary rotations $V,Y,Z$, and $W_i$.

We see that the circuit has a nice structure:
except the single prior rotation and the ancilla-controlled posterior rotations
on the system,
there is a sequence of multiplexers together with
ancilla-controlled rotations sandwiched in between those multiplexers.
This structure also holds for higher-dimensional cases,
which is straightforward to obtain from the multiple use of CSD.

\subsection{Quantum circuit of generalized extreme $(n,m)$-channels}
\label{sec:cirgextnm}

\subsubsection{Extreme qutrit-to-qubit and qubit-to-qutrit channels}

A generalized extreme qutrit-to-qubit channel, i.e. $(3,2)$-channel,
can be represented by three Kraus operators
$K_0$, $K_1$, $K_2 \in \mathcal{M}_{2,3}$,
and the dilated unitary operator $U\in SU(6)$ can be partitioned as
 \begin{equation}\label{eq:U32gext}
  U=\begin{pmatrix}
    K_0 & F_0  \\ K_1 & F_1\\ K_2 & F_2
  \end{pmatrix},
\end{equation}
whereas $F_0$, $F_1$, $F_2$ are matrices such that $U$ is unitary.
As each block is rectangular rather than square,
the initial and final unitary operators will have different sizes.
As the partition is $3\times 2$, we needs to employ CSD twice.
The first step leads to
$ U=A D B$ for
\begin{align}\label{} \nonumber
A&=\text{diag}(U_1,U_2), \\
B&=\text{diag}(V_1,V_2),
\end{align}
for $U_1\in SU(2)$, $U_2\in SU(4)$, $V_1,V_2\in SU(3)$,
and

\begin{align}\label{}
  D&=\left(\begin{array}{ccc|ccc}
  \cos\theta_1 & 0 & 0 & -\sin\theta_1 & 0 & 0 \\
  0 & \cos\theta_2 & 0 & 0 & -\sin\theta_2 & 0 \\ \hline
  0 & 0 & 0 & 0 & 0 & 1 \\
  \sin\theta_1 & 0 & 0 & \cos\theta_1& 0 & 0 \\ \hline
  0 & \sin\theta_2 & 0 & 0 & \cos\theta_2 & 0 \\
   0 & 0 & 1 & 0 & 0 & 0
  \end{array} \right) \\
  & \equiv \begin{pmatrix}
    D_{11} & D_{12} \\ D_{21} & D_{22} \\ D_{31} & D_{32}
  \end{pmatrix}.
\end{align}

\noindent The second step of CSD of $U_2$ leads to
$A=WEF$ for $W=\text{diag}(W_1,W_2,W_3)$,
$F=\text{diag}(\mathds{1},F_1,F_2)$,
$E=\text{diag}(\mathds{1},M)$,
for $W_1,W_2,W_3\in SU(2)$, for $F_1,F_2\in SU(2)$,
and $M$ is a qubit multiplexer
$M=\begin{pmatrix} C & -S \\ S & C\end{pmatrix}$
such that $C^2+S^2=\mathds{1}$.

We find the three Kraus operators are represented as
\begin{align}\label{eq:K32gext} \nonumber
  W_1 D_{11} V& = K_0, \\ \nonumber
  W_2 (CF_1D_{21}-SF_2D_{31}) V& = K_1, \\
  W_3 (SF_1D_{21}+CF_2D_{31})V& = K_2.
\end{align}

The unary representation of quantum circuit of qutrit-to-qubit generalized extreme channel is
\[  \includegraphics[width=.35\textwidth]{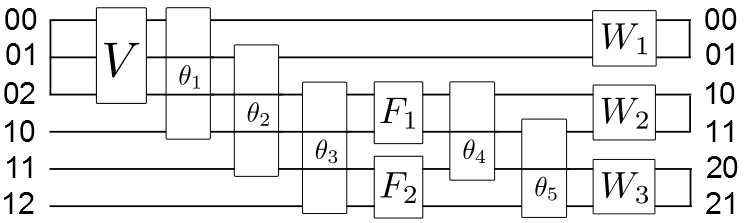}\]

\noindent Here a box with $\theta$ inside represents a Givens rotation of angle $\theta$,
and note $\theta_3=\pi/2$.
The input include a qutrit system and a qubit ancilla,
while the output include a qubit system and a qutrit ancilla.
The vertical lines at the right end represent trace operation,
and the output of the circuit is the sum of the three parts at the right end,
leading to a qubit output state.
Also, the vertical line at the left end represents the input state subspace.

We see that it is quite direct to represent the trace operation in the unary circuit.
Also, note that the encoding of the basis states is different for the input and output spaces,
with $\textsc{a}\otimes\textsc{s}$ for the input and
$\textsc{a}'\otimes\textsc{s}'$ for the output.
On the other hand, it is not so straightforward to draw the circuit diagram
with tensor-product structure due to the final trace operation.

Next we study the generalized extreme qubit-to-qutrit, i.e. $(2,3)$-channels,
with two Kraus operators.
The dilated unitary operator is also six dimension and partitioned as
 \begin{equation}\label{eq:Uq23gext}
  U=\begin{pmatrix}
    K_0 & F_0 & E_0 \\ K_1 & F_1 & E_1
  \end{pmatrix}.
\end{equation}
It is straightforward to see that this unitary operator can be viewed as
the inverse of the dilated unitary operator for generalized extreme $(3,2)$-channels.
After CSD,
we find the two Kraus operators are
\begin{align}\label{K23gext}
   K_0 = V_1 D_{11}^t W_1, \;      K_1 = V_2 D_{12}^t W_1.
\end{align}

Ignoring the initial and final rotations,
the quantum circuit for a generalized extreme $(2,3)$-channel is
just the inverse of the quantum circuit for a generalized extreme $(3,2)$-channel.
That is, if we run the circuit backward from right to left,
the input state is on the right hand side, and the output state is on the left hand side.
However, in this case there is only one rotation $W$
while two rotations $V_1$ and $V_2$.
For clarity, the circuit is shown below
\[
  \includegraphics[width=.35\textwidth]{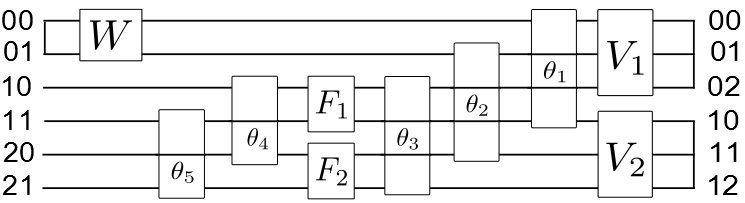}\]

\subsubsection{Extreme two qubit-to-qubit and qubit-to-two qubit channel}

The analysis to obtain the quantum circuits for generalized extreme
$(2,4)$- and $(4,2)$- channels
is similar with that for the previous cases.
For a generalized extreme $(4,2)$-channel, the dilated unitary operator can be partitioned as
\begin{equation}\label{eq:U42gext}
  U=\begin{pmatrix}
    U_{11} & U_{12}  \\ U_{21} & U_{22}
  \end{pmatrix},
\end{equation}
for $U\in SU(8)$, $U_{ij}\in \mathcal{M}_4$, ($i,j=1,2$)
and the four Kraus operators $K_i\in \mathcal{M}_{2,4}$ ($i=0,1,2,3$)
are contained in $U_{11}$ and $U_{21}$ such that
\begin{equation}\label{}
  U_{11}=\begin{pmatrix} K_0 \\ K_1  \end{pmatrix},
  U_{21}=\begin{pmatrix} K_2 \\ K_3  \end{pmatrix}.
\end{equation}

Employing CSD twice we find $K_i=W_i(K_{i0}, K_{i1})V$
for
\begin{align}\label{}\nonumber
    K_{00} & =C_1 F_1 C_2, \; K_{01} =-S_1 F_2 C_3, \\ \nonumber
    K_{10} & =S_1 F_1 C_2, \; K_{11} =C_1 F_2 C_3, \\ \nonumber
    K_{20} & =C_0 F_3 S_2, \; K_{21} =-S_0 F_4 S_3, \\
    K_{30} & =S_0 F_3 S_2, \; K_{31} =C_0 F_4 S_3,
\end{align}
with $F_i\in SU(2)$,
$C_i^2+S_i^2=\mathds{1}_2$,
$W_i\in SU(2)$, $V\in SU(4)$.

The unary circuit for a generalized extreme $(4,2)$-channel is
\[\includegraphics[width=.35\textwidth]{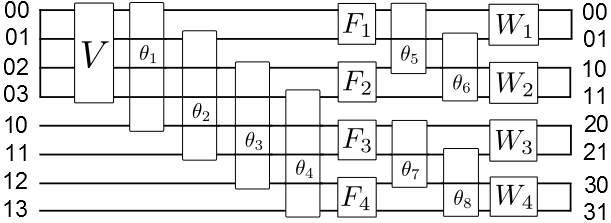}\]

For $(2,4)$-case the two Kraus operators are
\begin{equation}\label{K24gext}
  K_0=W_0\begin{pmatrix} C_2 F_1 C_1 \\ -C_3 F_2 S_1  \end{pmatrix}V,
  K_1=W_1\begin{pmatrix} -S_2 F_1 C_1 \\ S_3 F_2 S_1  \end{pmatrix}V.
\end{equation}
with $W_i\in SU(4)$, $V\in SU(2)$.
The quantum circuit is also straightforward to obtain,
so we would not show that here.

\subsection{Circuit costs}

After the explanation of the circuit design above,
here we analyze the number of primary gates in the quantum circuit.
For qudit case, there are $(d-1)$ multiplexers, each containing $d$ parameters.
There are $(d-2)$ controlled-rotations with the ancilla as controller,
and there are $(d+1)$ prior and posterior rotations.
Then the total parameter number is
\begin{equation}\label{}
  d(d-1)+(d^2-1)(d-2)+(d^2-1)(d+1)=(2d^2+2d-1)(d-1).
\end{equation}
However, the posterior rotations are classically controlled,
so only one posterior rotation contributes to the circuit cost.
Then we obtain the number of primary gates in the circuit is
\begin{equation}\label{eq:circostqud}
d(d-1)+(d^2-1)(d-2)+2(d^2-1)=d(d+2)(d-1)\in O(d^3).
\end{equation}
The order can also be obtained from a dimension counting argument.
In quantum circuit model,
the circuit costs for generating arbitrary quantum states or simulating unitary gates
are basically determined by the number of independent (real) parameters~\cite{SMB04,SPM+03,MVB04,BVM+05,BOB05,PB11}.
The dilated unitary operator for an arbitrary generalized extreme qudit channel
is of dimension $d^2$, which leads to $O(d^4)$ gates,
while the initial state of the ancilla is fixed as $|0\rangle$,
which would reduce the number of gates by order $d$,
eventually resulting in $O(d^3)$ gates.
Furthermore,
the order can also be obtained from the parameters of an arbitrary generalized extreme qudit channel.
It is well known that an $M$-dimensional while rank-$k$ hermitian matrix contains $k(2M-k)$ parameters
(also see Ref.~\cite{FL13}),
then a rank-$d$ generalized extreme qudit channel contains $d(2d^2-d)-d^2=2d^2(d-1)\in O(d^3)$ parameters,
with $d^2$ constraints from the trace preserving condition.

For the case of DA channels, a rank-$n$ generalized extreme channel in $\mathscr{S}_{n,m}$
contains $2n^2(m-1)\in O(n^2m)$ parameters,
with $n^2$ constraints from the trace preserving condition.
The dilated unitary operator for an arbitrary generalized extreme $(n,m)$-channel
is of dimension $mn$, which leads to $O(n^2m^2)$ gates.
With the constraint on the ancilla,
we obtain $O(n^2m)$ primary gates eventually in the circuit.
Quantum circuit lower bounds for simulating quantum channels
have been recently investigated using parameter counting~\cite{ICK+16,ICC16}
and different circuits have been designed,
which is consistent with our results here.

\section{Extreme channel circuit via ansatz}
\label{sec:ext2}

\begin{table*}[t!]
\begin{center}
\begin{tabular}{|c|c|c|c|c|c|c|}
\hline

\hline
\multirow{2}{*}{} & \multicolumn{2}{|c|}{\textsc{Ansatz I}} &  \multicolumn{2}{|c|}{\textsc{Ansatz II}} &  \multicolumn{2}{|c|}{\textsc{Ansatz III}} \\\cline{2-7}
 &                      \text{Parameter}  & \text{Precision} & \text{Parameter}  & \text{Precision} & \text{Parameter}  & \text{Precision}  \\\hline
$\mathcal{D}_2$                    & $23$  &$<10^{-4}$        &  $23$  &$<10^{-4}$       & $17$  &$<10^{-4}$        \\\hline
$\mathcal{D}_2\rightarrow \mathcal{D}_3$              & $65$  &$10^{-4}$         &  $65$  &$10^{-4}$      &  $41$  &$10^{-3}$   \\\hline
$\mathcal{D}_3\rightarrow \mathcal{D}_2$       & $55$  &$10^{-4}$        & $43$  &$10^{-3}$      & $31$  &$10^{-2}\sim10^{-3}$        \\\hline
$\mathcal{D}_3$   & $140$  &$10^{-3}$        & $116$  &$10^{-3}$       & $92$  &$10^{-2}$       \\\hline
$\mathcal{D}_2\rightarrow \mathcal{D}_4$              & $183$  &$10^{-3}$        &  $159$  &$10^{-3}$      &  $99$  &$10^{-3}$   \\\hline
$\mathcal{D}_4\rightarrow \mathcal{D}_2$       & $95$  &$10^{-3}$        & $71$  &$10^{-3}$      & $75$  &$10^{-2}$       \\\hline
$\mathcal{D}_4$   & $471$  &$10^{-2}$       & $351$  &$10^{-2}$       & $291$  &$10^{-1}$       \\\hline
\end{tabular}
\end{center}
\caption{\textbf{Optimization of quantum channel decomposition.}
The rows are for different channels,
and the columns are for different ansatz with the number of parameters in the optimization and simulation precision.}
\label{tab:simresult}
\end{table*}

Furthermore,
it is interesting to explore whether it is possible to achieve lower quantum circuit costs,
till the circuit lower bound $\Omega(d^2)$~\cite{BOB05},
which is the bound for a general qudit unitary operator,
for qudit channels and $\Omega(nm)$ for DA channels
while maintaining the accuracy of channel decomposition.
To achieve this, we need to employ decomposition methods other than CSD.

A quantum circuit ansatz has been proposed
for arbitrary extreme qudit channels and also generalized extreme channels,
which has circuit complexity $O(d^2)$~\cite{WS15}.
It relies on a construction of arbitrary extreme qudit channel
$\mathcal{E}^\text{e}\in\mathscr{S}_d$ with its representation $\{K_i\}$
such that
\begin{equation}
\label{eq:F}
	K_i:=W F_i V,\;
	F_i:=X_iE_i,\;
	E_i:=\sum_{j=0}^{d-1}a_{ij}Z_j,\,
	i\in\mathbb{Z}_d,
\end{equation}
for any $V, W \in SU(d)$,
$X_i =\sum_{\ell=0}^{d-1}|\ell\rangle\langle \ell+i|$ and
$Z_j =\sum_{\ell=0}^{d-1}\text{e}^{\text{i}2\pi \ell j/d}|\ell\rangle\langle \ell|$,
provided that $\{a_{ij}\in \mathbb{C}\}$
is chosen such that the set $\{F_i^\dagger F_j\}$
is linearly independent and $\sum_{i=0}^{d-1} F_i^\dagger F_i=\mathds{1}$
is satisfied.
The set $\{F_i^\dagger F_j\}$ forms a basis for $\mathcal{D}_d$,
with all $F_i^\dagger F_i$ diagonal
and $F_i^\dagger F_j$ ($i\neq j$) non-diagonal while one-sparse.
The unitary operators $V$ and $W$ serve to rotate the basis $\{F_i^\dagger F_j\}$
to other more general basis $\{K_i^\dagger K_j\}$.

For Kraus operators~(\ref{eq:F}), a quantum circuit ansatz is also proposed~\cite{WS15},
which is
\begin{equation}
\label{eq:extansatzIII}
	SU(d^2) \ni U:= \prod_{i=d-1}^1 C_i X_i \prod_{j=d-1}^1\prod_{k=j-1}^0 M_{jk}(\alpha_{jk},\beta_{jk}),
\end{equation}
for $C_iX_i:= X_i\otimes|i\rangle \langle i|$,
$C_jG_{jk}(\theta):=|j\rangle\langle j|\otimes G_{jk}(\theta)$,
$M_{jk}(\alpha,\beta):=C_jG_{jk}(\alpha)C_kG_{kj}(-\beta)$,
such that $F_i:=\langle i|U|0\rangle$ forms a linearly independent set $\{F_i^\dagger F_j\}$.
Note that here each gate $M_{jk}(\alpha_{jk},\beta_{jk})$ is a qubit-valued multiplexer,
i.e.\, it only contains two controlled-Givens rotations.
Without the sequence of $C_i X_i$, the circuit leads to a set of diagonal Kraus operators,
which will not form a linearly independent set in general.
The sequence of $C_i X_i$ serves to satisfy the linear independence condition.

Furthermore, as a modification of (\ref{eq:extansatzIII}),
here we show that it is also valid to have a circuit ansatz using multi-valued quantum multiplexers
instead of a sequence of qubit-valued multiplexers.
With the definition~(\ref{eq:quditMulti}),
the unitary operator $U\in SU(d^2)$ with
\begin{equation}
\label{eq:extansatzIII2}
	U:= \prod_{i=d-1}^1 C_i X_i \prod_{i=0}^{d-2} M_{i,i+1},
\end{equation}
leads to
$F_i:=\langle i|U|0\rangle$ that form a linearly independent set $\{F_i^\dagger F_j\}$
required by~(\ref{eq:F})
for most of the rotations angles contained in $M_{i,i+1}$.

Now we are at a stage to compare the method based on CSD in section~\ref{sec:ext}
and the construction~(\ref{eq:extansatzIII2}).
For clarity, we refer to the construction based on CSD as Ansatz I,
and we refer to~(\ref{eq:extansatzIII2}) as Ansatz III.
The reason for choosing those names is that we also find an intermediate formula, which is called Ansatz II.

In Ansatz III, the posterior rotations are the same for each Kraus operator,
and the linear independence is ensured by the sequence of $C_i X_i$ gates in~(\ref{eq:extansatzIII2}).
In Ansatz I, there is no such a sequence of $C_i X_i$ gates at the end of the circuit,
while the posterior rotations are different for each Kraus operator,
which ensures the linear independence condition.
Last, Ansatz II is motivated by the
so-called higher-order generalized singular value decomposition (HO GSVD)~\cite{PSV+11},
and is also a simplified version of Ansatz I.

For completeness, here we briefly describe the HO GSVD.
The original construction is for real matrices,
while can be straightforwardly applied to complex matrices.
Given a set $\{D_i\in \mathcal{M}_{m_i,n}\}$ of $N$ full column rank matrices $D_i$,
the HO GSVD is defined as $D_i=U_i \Sigma_i V$ $(i=1,\dots,N)$,
and $U_i$ contains the normalized left basis vectors of $D_i$,
$V$ is formed by the normalized right basis vectors determined by all $D_i$,
and each $\Sigma_i$ is a diagonal positive matrix
containing generalized singular values of each $D_i$.
A procedure to obtain $V$, $U_i$, and
$\Sigma_i$ has also been constructed~\cite{PSV+11}.

\begin{figure*}[t!]
  \centering
  \includegraphics[width=.4\textwidth]{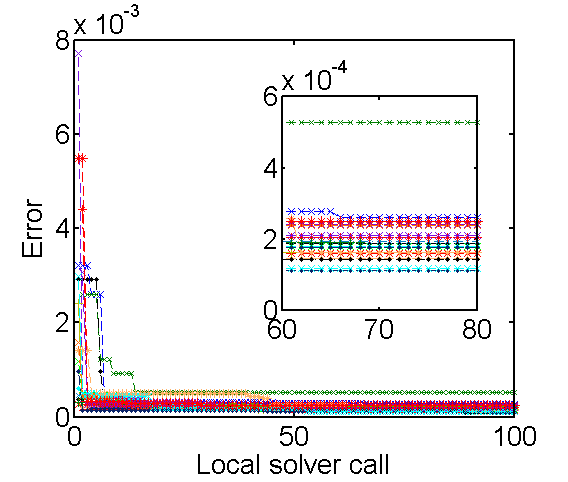}
  \includegraphics[width=.4\textwidth]{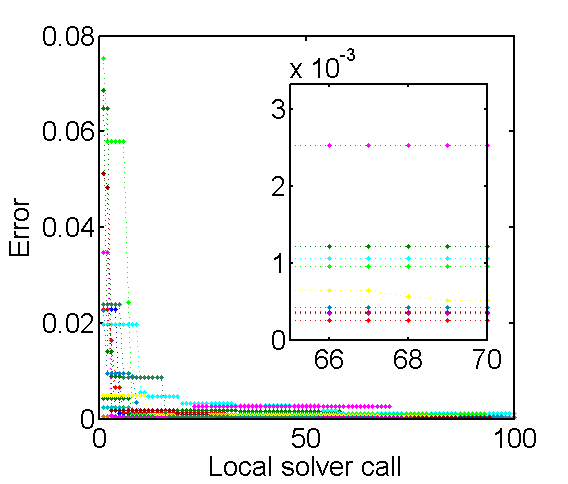}
  \caption{Simulation result for 20 randomly chosen qubit-to-qutrit channels.
  (Left) Ansatz I, which is the same with Ansatz II.
  (Right) Ansatz III.
  The number of local solver calls is proportional to the runtime of the simulation.
  The simulation error is in terms of trace distance on Choi state.
  }\label{fig:Q23}
\end{figure*}

In HO GSVD, the prior and posterior operators are not necessarily unitary.
We intend to modify it by substituting the prior and posterior operators by unitary operators
and then form an approximate representation of generalized extreme channels.
Given a set of Kraus operators $\{K_i\in \mathcal{M}_{m,n}\}$
of a random generalized extreme channel,
and if generically each $K_i$ is of full column rank,
then we employ the approximation
\begin{equation}\label{eq:approxK}
K_i\approx U_i \Sigma_i V
\end{equation}
for unitary matrices $U_i\in SU(m)$ and $V\in SU(n)$.
The trace-preserving condition specifies
$\sum_i \Sigma_i^2 =\mathds{1}$.
This is equivalent to simplify the Ansatz I by ignoring the ancilla-controlled rotations
and only keep the system-controlled rotations.
For instance,
the rotation $Z_1$ from Eq.~(\ref{eq:Kqutgext}) is ignored
for the qutrit case,
and the rotations $Z$ and $Y$ from Eq.~(\ref{eq:Kqubbgext}) are ignored
for the two-qubit case,
while for the qubit case
Ansatz II is the same with Ansatz I,
see Eq.~(\ref{eq:Kqubgext}).
This, of course, cannot be used to
represent arbitrary channels in general.
However, our simulation results show that
this can provide a relatively good and practical ansatz
for the approximate decomposition of channels up to dimension four,
demonstrated in Table~\ref{tab:simresult}.

In general, ignoring the prior and posterior rotations,
for Ansatz II the unitary circuit for generalized extreme qudit channel is
\begin{equation}\label{eq:extansatzII}
  U=\prod_{i=0}^{d-2} M_{i,i+1}=\prod_{i=0}^{d-2}\prod_{l=0}^{d-1} C_l G_{i,i+1}(\theta_l).
\end{equation}
The formulas for generalized extreme $(n,m)$-channels in Ansatz II
can also be obtained.

For clarity, we now have three different circuit constructions:
I) The one based on CSD, named as Ansatz I;
II) The simplified CSD, or the modified HO GSVD, named as Ansatz II;
III) The method in~\cite{WS15} and modified as in Eq.(\ref{eq:extansatzIII2}), named as Ansatz III.
Ansatz II is obtained from Ansatz I by ignoring the ancilla-controlled rotations,
and Ansatz III is obtained from Ansatz II by setting the posterior rotations same for each Kraus operator,
and adding a sequence of ancilla-controlled $X_i$ gates.

The reason to consider different circuits other than the one based on CSD
is to reduce the circuit cost
especially for the benefit of practical implementation.
We find that Ansatz II and III have circuit cost $O(d^2)$,
instead of $O(d^3)$ for qudit channels,
and $O(nm)$ for DA channels instead of $O(n^2m)$.
This means that the circuit lower bound can be achieved
without affecting the accuracy of the circuit,
as demonstrated below by the numerical simulations in section~\ref{sec:decom}.
Note that there is no conflict between our circuit costs
in Ansatz II and III and the circuit lower bound, which is achieved by Ansatz I,
since they are used for {\em approximate} instead of exact channel decomposition.

\section{Quantum channel decomposition}
\label{sec:decom}

It is conjectured by Ruskai~\cite{Rus07} that any channel $\mathcal E \in\mathscr{S}_{n,m}$
can be decomposed as
\begin{equation}\label{eq:dec}
  \mathcal E=\sum_{i=1}^m p_i \mathcal{E}^\text{g}_i,
\end{equation}
for $\mathcal{E}^\text{g}_i \in \mathscr{S}_{n,m}^{\leq n}$,
and probability $p_i\in [0,1]$ such that $\sum_{i=1}^m p_i=1$.
The case for $n=m=2$ has been proved~\cite{RSW02},
and the case for $m=2$ is also proved~\cite{Rus07} by extending the method in Ref.~\cite{RSW02}.
The cases for $n=m=3,4$ has been numerically verified (with errors)
using a quantum circuit for generalized extreme channels~\cite{WS15},
which is equivalent to our Ansatz III here.

Channel partition into a sum of other smaller channels always exists,
while the problem is that the number of partitions may be bigger than $m$,
and an analytical formula or algorithmic procedure for such a decomposition is not known.
Despite this obstacle,
in this work we rely on optimization for such decompositions,
and our simulation for dimensions up to four yields positive results for
this channel decomposition conjecture.

The distance between two quantum channels~$\mathcal E$ and $\tilde{\mathcal E}$ is defined as
the diamond-norm distance~\cite{AKN97}
\begin{align}
\label{eq:diamond}
	\|\mathcal{E}-\mathcal{\tilde{E}}\|_\diamond
		:=\|\mathcal{E}\otimes\mathds{1}-\mathcal{\tilde{E}}
			\otimes\mathds{1}\|_{1\rightarrow 1},
\end{align}
for the induced Schatten one-norm
\begin{equation}
	\|\mathcal{E}-\mathcal{\tilde{E}}\|_{1\rightarrow 1}
		:=\max_\rho\|\mathcal{E}(\rho)-\mathcal{\tilde{E}}(\rho)\|_1,
\end{equation}
and for $\mathds{1}$ acting on an ancillary space.
The diamond-norm distance characterizes the operational distance
between two channels,
similar with the trace distance on quantum states.
That is, the success probability for correctly
distinguishing two channels from their output states,
based on trace distance, is
$\frac{1+\epsilon/2}{2}$ for diamond distance $\epsilon$.
However, the diamond distance is hard to compute,
so following~\cite{WS15},
we employ Choi states and the trace distance~$D_t(\mathcal{C}, \mathcal{\tilde{C}})$
since~\cite{Wat13}
\begin{equation}\label{eq:cbound}
	\|\mathcal{E}-\mathcal{\tilde{E}}\|_{\diamond}\leq 2n D_t(\mathcal{C}, \mathcal{\tilde{C}}).
\end{equation}

We have performed numerical simulation using the three ansatz
on Matlab using GlobalSearch algorithm.
The number of parameters in the optimization
and the simulation precision is shown in Table~\ref{tab:simresult}.

For instance, for qubit-to-qutrit channels,
we employ the decomposition
\begin{equation}\label{eq:dec32}
  \mathcal E= p_1 \mathcal{E}^\text{g}_1 + p_2 \mathcal{E}^\text{g}_2+ p_3 \mathcal{E}^\text{g}_3,
\end{equation}
for $\mathcal{E}^\text{g}_i \in \mathscr{S}_{2,3}^{\leq 2}$,
and probability $p_i$.
The simulation result for 20 randomly chosen qubit-to-qutrit channels
using different ansatz are shown in Fig.~\ref{fig:Q23}.
We can see that Ansatz I, which is the same with Ansatz II for this case,
produces simulation accuracy one order better than Ansatz III.

In general, from Table~\ref{tab:simresult}
we find that Ansatz II can lead to the same simulation precision with Ansatz I,
and Ansatz II leads to precision one order better than Ansatz III.
We conclude that for approximate quantum channel decomposition,
Ansatz II is our best choice since the quantum circuit cost
is lower than that for Ansatz I.

\section{Discussion and conclusion}
\label{sec:conc}

In this work we have investigated quantum channel decomposition problem
in terms of convex sum of extreme channels.
Ruskai's quantum channel decomposition conjecture~\cite{Rus07}
is numerically tested here for low-dimensional cases.
The quantum circuit representation of generalized extreme channels
may provide hints and lead to an approach for {\em exact} quantum channel decomposition.
Furthermore, for quantum channel simulation purpose
our work extends the approach~\cite{WS15} to the cases of dimension-altering channels
and improves the quantum circuit design.
Quantum channel simulation using extreme channel decomposition is deterministic and resource-optimal,
and serves as a unique approach for quantum simulation.

Our quantum circuits used in this work
are constructed {\em a priori} based on cosine-sine decomposition,
hence {\em Ansatz}, which means it is not known
how to exactly decompose a generalized extreme channel into our quantum circuit.
The ansatz approach is in particular suitable for the use in optimization algorithms.
Three different ansatz for generalized extreme channel circuits
are proposed, and the best one, Ansatz II,
demonstrates high simulation precision with relatively small circuit costs.
Although Ansatz II can achieve
the cost $O(d^2)$,
the more accurate Ansatz I can only achieve $O(d^3)$,
which follows from parameter counting.
Furthermore, Ansatz II is also motivated by the HO GSVD,
which has demonstrated broad applications in field of matrix analysis,
while its validity for higher than four dimensional channels remains to be investigated.

Our circuit construction employs Givens rotation and its controlled version,
while other universal set of gates also exist,
such as Householder reflections.
It remains to investigate whether better circuit design can be achieved
using other set of gates and other methods.
We note that alternative circuit designs have been studied recently~\cite{ICK+16,ICC16},
which shows that classical control can further reduce the circuit costs.
Comparison of circuit designs could be made while this is beyond the scope of our current work.

In addition, the classical optimization plays nontrivial roles,
hence it is valuable to explore alternative optimization routines.
A slight variation of the optimization employed here,
which is ``global,''
is to optimize locally each generalized extreme channel
while keeping the others fixed.
Such a step-by-step local optimization is common in many optimization algorithms,
and thus can be employed for the channel decomposition problem.

\begin{acknowledgments}
Funding support from NSERC of Canada and
a research fellowship at Department of Physics and Astronomy, University of British Columbia
are acknowledged.
\end{acknowledgments}

\bibliography{ext}
\end{document}